\documentclass[twocolumn,aps,prb,superscriptaddress,amsmath,amssymb,superscriptaddress]{revtex4-1}
\newcommand{\pagenumbaa}{1}
\usepackage{graphicx}

\usepackage{verbatim}
\usepackage{amsmath}
\usepackage{float}
\usepackage{hyperref}
\usepackage{mathtools}
\DeclarePairedDelimiter\abs{\lvert}{\rvert}
\newcommand{\fakesection}[1]{
  \par\refstepcounter{section}
  \sectionmark{#1}
  \addcontentsline{toc}{section}{\protect\numberline{\thesection}#1}
}

\begin{document}

\title{Correlations between Optical Properties and Voronoi-Cell Area of Quantum Dots}

\author{Matthias C. L\"obl}
\email{matthias.loebl@unibas.ch}
\affiliation{Department of Physics, University of Basel, Klingelbergstrasse 82, CH-4056 Basel, Switzerland}
\author{Liang Zhai}
\affiliation{Department of Physics, University of Basel, Klingelbergstrasse 82, CH-4056 Basel, Switzerland}
\author{Jan-Philipp Jahn}
\affiliation{Department of Physics, University of Basel, Klingelbergstrasse 82, CH-4056 Basel, Switzerland}
\author{Julian Ritzmann}
\affiliation{Lehrstuhl fur Angewandte Festk\"orperphysik, Ruhr-Universit\"at Bochum, D-44780 Bochum, Germany}
\author{Yongheng Huo}
\affiliation{\makebox[0.8\textwidth][s]{Institute for Integrative Nanosciences, IFW Dresden, Helmholtzstrasse 20, D-01069 Dresden, Germany}}
\affiliation{Hefei National Laboratory for Physical Sciences at the Microscale (HFNL) \& Shanghai Branch of the University of Science and Technology of China (USTC), No.99, Xiupu Road, Pudong New District 201315, Shanghai, China}
\author{Andreas D. Wieck}
\affiliation{Lehrstuhl fur Angewandte Festk\"orperphysik, Ruhr-Universit\"at Bochum, D-44780 Bochum, Germany}
\author{Oliver G. Schmidt}
\affiliation{\makebox[0.8\textwidth][s]{Institute for Integrative Nanosciences, IFW Dresden, Helmholtzstrasse 20, D-01069 Dresden, Germany}}
\author{Arne Ludwig}
\affiliation{Lehrstuhl fur Angewandte Festk\"orperphysik, Ruhr-Universit\"at Bochum, D-44780 Bochum, Germany}
\author{Armando Rastelli}
\affiliation{Institute of Semiconductor and Solid State Physics, Johannes Kepler University Linz, Altenbergerstrasse 69, A-4040 Linz, Austria}
\author{Richard J. Warburton}
\affiliation{Department of Physics, University of Basel, Klingelbergstrasse 82, CH-4056 Basel, Switzerland}

\begin{abstract}
A semiconductor quantum dot (QD) can generate highly indistinguishable single-photons at a high rate. For application in quantum communication and integration in hybrid systems, control of the QD optical properties is essential. Understanding the connection between the optical properties of a QD and the growth process is therefore important. Here, we show for GaAs QDs, grown by infilling droplet-etched nano-holes, that the emission wavelength, the neutral-to-charged exciton splitting, and the diamagnetic shift are strongly correlated with the capture zone-area, an important concept from nucleation theory. We show that the capture-zone model applies to the growth of this system even in the limit of a low QD-density in which atoms diffuse over $\mu$m-distances. The strong correlations between the various QD parameters facilitate preselection of QDs for applications with specific requirements on the QD properties; they also suggest that a spectrally narrowed QD distribution will result if QD growth on a regular lattice can be achieved.
\end{abstract}

\maketitle

\setcounter{page}{\pagenumbaa}
\thispagestyle{plain}

\fakesection{Introduction}
\label{sec:introduction}

Semiconductor quantum dots (QDs) are excellent as a bright source of highly indistinguishable single photons \cite{Huber2017,Ding2016,Somaschi2016,Kirsanske2017,Liu2018,He2017,Gschrey2015,Scholl2019} and entangled photon pairs \cite{Stevenson2006,Akopian2006,Hudson2007,Chen2018,Huber2018,Liu2019,Basset2019,Wang2019}. A QD can host a single spin \cite{Atature2006,Bechtold2015,Javadi2017,Lobl2017} which, however, has a too short coherence time for applications in quantum communication \cite{Prechtel2016,Delteil2016,Majcher2017,Ding2019}. A hybrid-system of a QD and an atomic quantum memory is more promising in that respect \cite{Sangouard2007,Rakher2013}. To connect a QD to an atomic memory based on rubidium, the QD should emit photons matched both in emission energy and bandwidth to the memory \cite{Wolters2017}. The emission energy can be matched by using GaAs QDs embedded in AlGaAs \cite{Jahn2015,Keil2017}; bandwidth matching can be achieved by using a Raman-scheme \cite{Beguin2018,Pursley2018,Lee2018,He2013}.

The growth of QDs has been intensively studied employing scanning probe microscopy \cite{Rastelli2004,Bruls2002,Atkinson2012,Fanfoni2012,Heyn2007,Rastelli2008,Nothern2012,Miyamoto2009,Bietti2014}. Aiming at entangled photon-pair generation, a connection between such an analysis and the optical properties \cite{Rastelli2004b,Surrente2017,Sapienza2017} has focused mostly on the fine-structure splitting of the QD-emission \cite{Huo2013,Plumhof2010,Abbarchi2010,Mereni2012,Huo2014,Liu2014,Skiba2017,Basset2018}. To tailor all the optical QD-properties, it is important to understand how they are connected to the QD-growth \citep{Atkinson2012}.

In this Letter, we establish a strong correlation between the optical properties of GaAs QDs, such as emission energy and diamagnetic shift, and a basic concept from nucleation theory, the capture-zone \cite{Mulheran1996,Blackman1996,Miyamoto2009}. This correlation is not obvious since the applicability of the capture zone model depends on conditions such as sudden nucleation \cite{Fanfoni2007} and spatially uniform diffusion \cite{Ratto2006}. Both are not necessarily fulfilled. The correlations that we find here are absent for InGaAs QDs \cite{Spencer2019} and locally very weak for SiGe QDs \cite{Miyamoto2009,Ratto2006}. The system we investigate consists of GaAs QDs grown by infilling of Al-droplet-etched nano-holes in an AlGaAs surface\cite{Wang2007,Huo2013}.

We apply the capture-zone model to the first phase of this process, the formation and growth of Al-droplets. The capture-zone model implies a correlation between Al-droplet volume (island size \cite{Pratontep2004,Miyamoto2009}) and capture-zone area. We show experimentally that this results in a strong correlation between the capture zone area and the optical QD-properties.

We use spatially-resolved photoluminescence imaging to determine simultaneously the position and the optical properties of individual QDs. We investigate samples of low QD-density where a QD is small relative to the distances between QDs (point-island model \cite{Blackman1996}). The capture zone of each QD is determined as its Voronoi-cell (VC) \cite{Mulheran1996,Miyamoto2009,Fanfoni2007,Ratto2006}. From the distribution of the VC-areas, we estimate a critical nucleus size of the Al-droplets (supplemental material). We find a strong negative correlation between the VC-area, $A_{\text{VC}}$, of a QD and its emission energy, in turn, a strong positive correlation between $A_{\text{VC}}$ and the diamagnetic shift of the emission. These results can be well explained with the capture-zone model describing the aggregation dynamics of the Al-droplets. Correlations are measured for QDs of particularly low density, $n_{\text{QD}}<1\ \mu\text{m}^{-2}$, implying a diffusion length on the $\mu$m-scale during the growth phase of Al-droplets \cite{Bietti2014}. This is a striking result: The optical properties of a QD, which are directly related to its structure on an nm-scale, are strongly correlated with its surroundings on a $\mu$m-scale.

\begin{figure*}[t!]
\includegraphics[width=1.8\columnwidth]{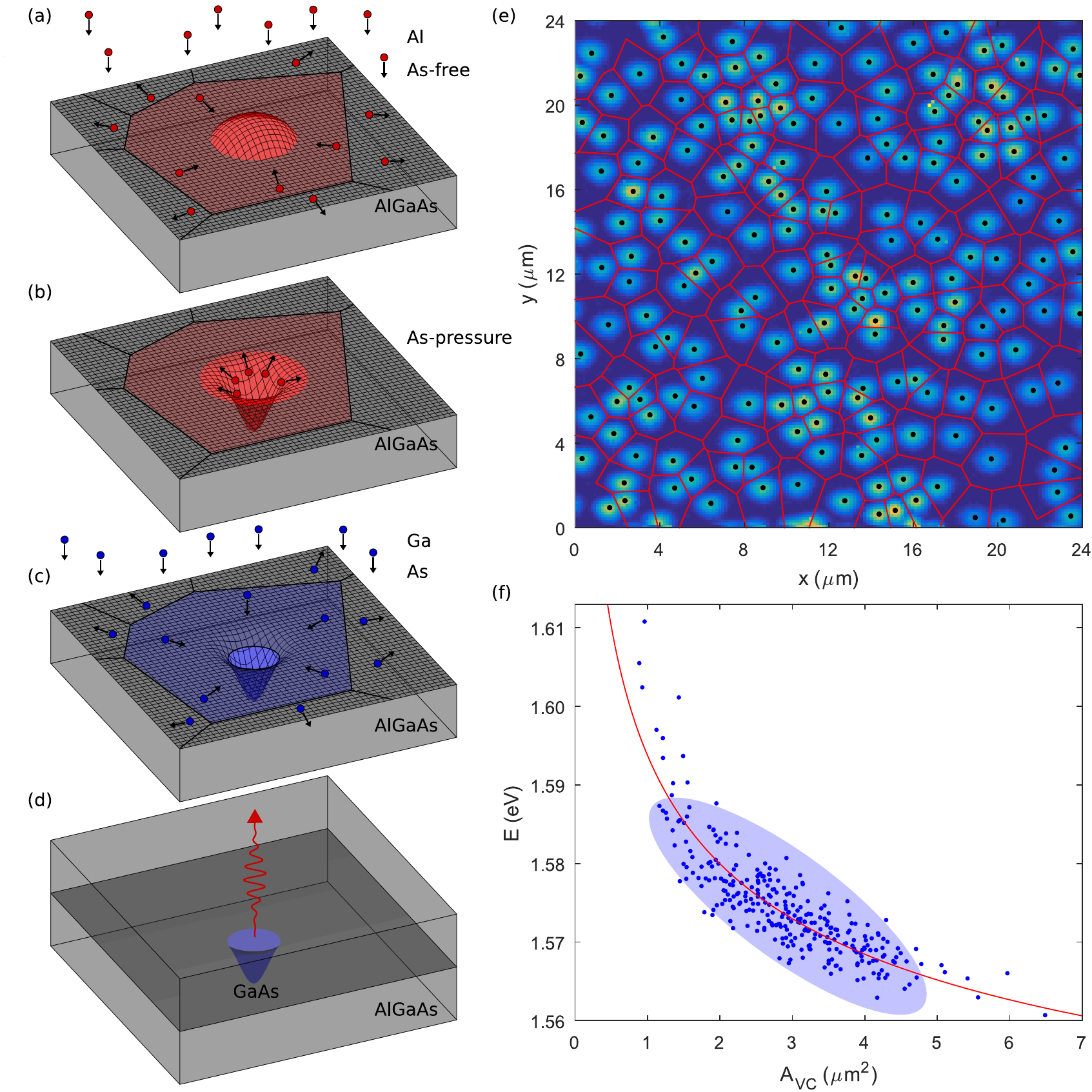}
\caption{(a) Schematic of the growth process of the quantum dots (QDs). In a first step, aluminum is deposited on an epitaxially grown AlGaAs surface. The aluminum atoms nucleate in the form of liquid nano-droplets. An atom is most likely to attach to the closest Al-droplet. This is the Al-droplet into whose capture zone (red area) the atom falls. (b) Underneath the Al-droplet, the substrate material is unstable and nano-hole etching takes place upon exposure to an arsenic flux. (c) After formation of nano-holes, GaAs is deposited. Diffusion leads to an infilling of the droplet-etched nano-holes with GaAs. (d) Finally, the sample is capped with AlGaAs. The GaAs within the nano-hole is now embedded in higher bandgap AlGaAs and forms a QD. (e) Spatially resolved photoluminescence (PL) on a $24\times24\ \mu\text{m}^2$-large region of sample A. Positions of individual QDs are obtained by Gaussian fitting (black dots). The red lines are the Voronoi-cells (VCs) corresponding to the QD-positions. (f) Relation between the VC-area ($A_{\text{VC}}$) and the emission energy of the neutral exciton, $X^0$. The light-blue ellipse is a guide to the eye indicating the correlation in a linear approximation (correlation coefficient $\rho=-0.812$). The red line is a fit of Eq.\ \ref{Eq:E_AVC}.}
\label{fig:VC}
\end{figure*}

\textit{Sample growth} -- The samples are grown by molecular-beam epitaxy (MBE) on a (001)-substrate. We investigate two different samples (denoted here as A and B) that are grown in two different MBE-chambers. QDs are grown by GaAs-infilling of Al-droplet-etched nano-holes. A schematic depiction of the growth is shown in Fig.\ \ref{fig:VC}. First, aluminum is deposited on an AlGaAs-surface in the absence of an As-flux. The growth parameters are: 0.5 ML on Al$_{0.4}$Ga$_{0.6}$As, $T=600^{\circ}$C, flux $F=0.5$ ML/s for sample A; and 0.16 ML on Al$_{0.33}$Ga$_{0.67}$As, $T=630^{\circ}$C, flux $F=0.18$ ML/s for sample B. For both samples, the layer thicknesses correspond to an equivalent amount of AlAs. The Al-atoms nucleate (Volmer-Weber mode \cite{Heyn2015}) in the form of liquid nano-droplets on the sample surface (see Fig.\ \ref{fig:VC}(a)).

Underneath an Al-droplet the substrate material is unstable leading to a nano-etching process (Fig.\ \ref{fig:VC}(b)) \cite{Wang2007}. Under a low As-flux, the etching proceeds until the whole Al-droplet is consumed and the material is recrystallized in the surrounding region. Another $2$ nm of GaAs are grown on top, filling up the nano-holes via diffusion during a $2$-minute annealing step (Fig.\ \ref{fig:VC}(c)). Overgrown with AlGaAs, the filled nano-holes become optically active QDs (Fig.\ \ref{fig:VC}(d)).

\begin{figure*}[t!]
\includegraphics[width=2\columnwidth]{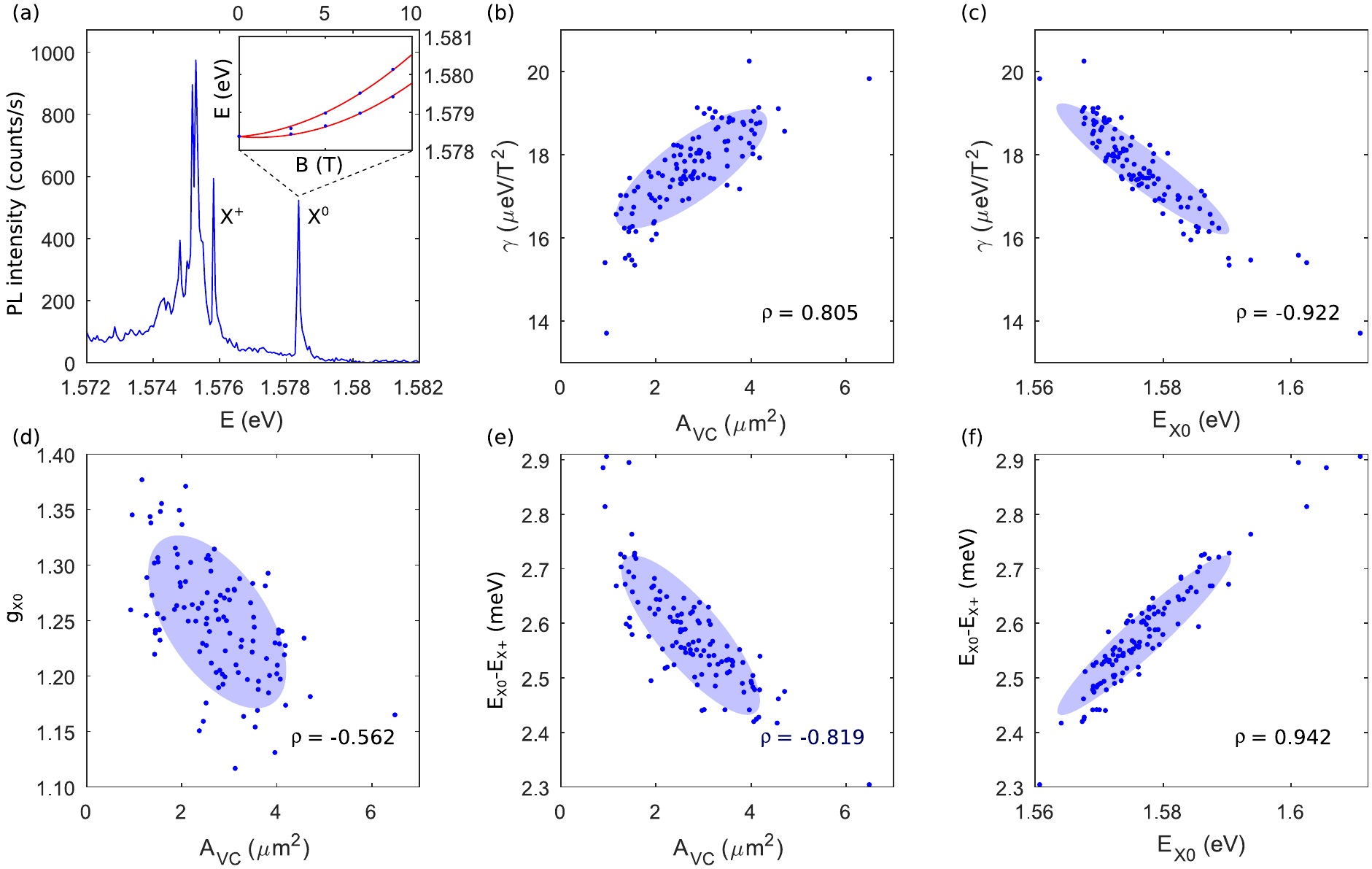}
\caption{(a) An exemplary PL-spectrum of a QD. The emission line at the highest energy is the neutral exciton ($X^0$). At lower energy, emission of a singly-charged exciton ($X^+$) and a broad emission from further excitons appear. The inset shows the $X^0$ emission energy as a function of the magnetic field. A diamagnetic shift and a Zeeman splitting are observed. The following sub-figures refer to emission from $X^0$ on sample A. (b) Diamagnetic shift as a function of $A_{\text{VC}}$ ($\rho=0.805$). The light blue ellipse has the same slope as a linear fit to the data points; its widths indicate a $1.5\sigma$-interval parallel and perpendicular to the slope. (c) Diamagnetic shift as a function of the emission energy of the neutral exciton. These parameters are negatively correlated (correlation coefficient $\rho = -0.922$ ). (d) A weak correlation between the exciton $g$-factor and the Voronoi-cell area, $A_{\text{VC}}$. (e) Splitting between neutral and positively-charged excitons ($E_{X0}-E_{X+}$) as a function of $A_{\text{VC}}$ ($\rho=-0.819$). (f) $E_{X0}-E_{X+}$ as a function of $X^0$ emission energy ($\rho = 0.942$).}
\label{fig:corr}
\end{figure*}

\textit{Experiment} -- Optical measurements are performed in a helium bath-cryostat. Photoluminescence (PL) is measured under above-band excitation ($\lambda=633\ \text{nm}$). An aspheric objective lens ($\text{NA}=0.68$) collects the PL. The PL of the QD-ensemble is centered at wavelength $787.4\ \text{nm}$ for sample A, $798.0\ \text{nm}$ for sample B (values referring to the neutral exciton, $X^0$), with ensemble standard-deviations $3.4\ \text{nm}$ and $1.4\ \text{nm}$, respectively. A typical spectrum of a single QD is shown in Fig.\ \ref{fig:corr}(a). The neutral exciton has the highest emission energy \cite{Jahn2015}; the positively charged exciton is redshifted by on average $2.7$ meV (sample B: $2.1$ meV). Additional exciton complexes appear at even lower energy \cite{Wang2009} but are not the focus of our analysis.

Spatially resolved micro-PL is performed by scanning the sample with a low-temperature piezoelectric xy-scanner (attocube ANSxy100/lr). A spatially resolved PL-measurement is shown in Fig.\ \ref{fig:VC}(e). QDs can be identified as bright regions on this PL-map. The lateral positions of the QDs are obtained by fitting two-dimensional Gaussians. A slight non-linearity of the piezo-scanner is corrected by using the widths of the fitted Gaussians as a reference (supplemental material). We determine the capture zone around a QD by its Voronoi-cell (VC) -- the area that is closer to this particular QD than to any other one. A Voronoi-diagram together with the corresponding QD-positions is shown in Fig.\ \ref{fig:VC}(e). We find an average VC-area of $\langle A_{\text{VC}}\rangle=3.04\pm0.08\ \mu\text{m}^2$ (sample A) and $\langle A_{\text{VC}}\rangle=6.87\pm0.44\ \mu\text{m}^2$ (sample B), corresponding to a quantum dot density of $n_{\text{QD}}=0.329\pm0.009\ \mu\text{m}^{-2}$ and $n_{\text{QD}}=0.146\pm0.009\ \mu\text{m}^{-2}$, respectively.

Shown in Fig.\ \ref{fig:VC}(f) is the emission energy of the neutral exciton, $E_{X^0}$, for many QDs as a function of the VC-area, $A_{\text{VC}}$. The plot is obtained by combining three independent PL-maps from sample A. We find a strong negative (Pearson) correlation coefficient of $\rho=-0.812$ (sample B: $\rho=-0.809$) between emission energy and $A_{\text{VC}}$ ($\rho=\pm1$ maximum correlation; $\rho=0$ no correlation). We explain this correlation by applying the capture-zone model to the growth phase of the Al-droplets (Fig.\ \ref{fig:VC}(a)). An Al-atom, impinging at a random position on the sample, moves on the surface via diffusion and is most likely to attach to the closest Al-droplet. In the capture-zone model, the growth rate of an Al-droplet is thus assumed to be proportional to the VC-area. If all Al-droplets form at about the same time (sudden nucleation \cite{Fanfoni2007}), it leads to a correlation between Al-droplet volume and VC-area. For an Al-droplet with a small VC-area, much material nucleates at its nearest neighbors reducing its own accumulation rate. In turn, an Al-droplet with a larger VC-area accumulates more atoms and the droplet-volume $V_d$ becomes larger. For a larger Al-droplet, the nano-hole etched underneath it becomes deeper \citep{Atkinson2012}. The QD subsequently formed from this nano-hole has a weaker confinement in the growth direction lowering its emission energy.

We obtain a quantitative relation between $A_{\text{VC}}$ and the emission energy by the following considerations. In the capture-zone model, the volume of each Al-droplet is proportional to $A_{\text{VC}}$. We assume that all Al-droplets have a similar aspect ratio \cite{Huo2014,Masafumi2013}. Then the droplet-height $H_d$ is connected to the droplet-volume $V_d$ and to the VC-area by $H_d\propto V_d^{1/3} \propto A_{\text{VC}}^{1/3}$. For the relation between QD-height $H$ (nano-hole depth, respectively) and the droplet-height $H_d$ \cite{Atkinson2012}, we assume a phenomenological relation $H\propto H_d^{\beta}$. Since $H$ is much smaller than the lateral extent of a QD \cite{Atkinson2012}, it is this parameter which mainly determines the emission energy of the QD. In the case of a hard-wall confinement and without considering Coulomb interaction terms, the emission energy of a QD is given by the bandgap plus electron and hole confinement energy: $E=E_0+\frac{h^2}{8\mu H^2}$, where $E_0=1.519\ \text{eV}$ is the bandgap of the QD-material (GaAs) and $\mu=\left(\frac{1}{m_e^*}+\frac{1}{m_h^*}\right)^{-1}$ the reduced electron-hole effective mass ($m_e^*=0.067m_e$, $m_h^*=0.51m_e$). Including the $d=2\ \text{nm}$ thick quantum well above the QDs leads to $E=E_0+\frac{h^2}{8\mu (H+d)^2}$. Using the above relations one obtains an equation connecting $A_{\text{VC}}$ and the emission energy $E$ of the QD:
\begin{equation}
\label{Eq:E_AVC}
E = E_0 + \frac{h^2}{8\mu}\cdot\left(\left(\alpha\cdot A_{\text{VC}}\right)^{\beta/3}+d\right)^{-2}
\end{equation}
A fit of Eq.\ \ref{Eq:E_AVC} to the data is shown in Fig.\ \ref{fig:VC}(f). With the fit parameters $\alpha=1.268\cdot10^{-32}\ \text{m}^{3/\beta-2}$ and $\beta=0.556$, we find a very good agreement with the data. The average height of a QD resulting from this fit is $H=8.7\ \text{nm}$ which agrees well with AFM-measurements (supplemental material). A direct correlation between the measured emission energy $E_{X^0}$ and the term on the right-hand side of Eq.\ \ref{Eq:E_AVC} shows an even higher correlation of $\rho=0.879$ (supplemental material) than that between emission energy and $A_{\text{VC}}$. This strong correlation supports our model for the connection between QD-properties and capture zone.

We consider further QD-properties and their connection to the VC-area, $A_{\text{VC}}$: the diamagnetic shift of the QD-emission; and the splitting between neutral and positively-charged exciton, $E_{X^0}-E_{X^+}$. Both of these QD-properties are connected mainly to the lateral rather than the vertical confinement of the QD.

We measure the energy of the PL-emission ($X^0$) as a function of a magnetic field applied in the growth direction (inset to Fig.\ \ref{fig:corr}(a)). The magnetic field splits the emission lines by the Zeeman energy and leads to a diamagnetic shift. For every QD we fit the relation \cite{VanHattem2013,Tholen2016}
\begin{equation}
\label{Eq:diamag}
E\left(B\right)=E(B=0)+\gamma B^2\pm\frac{1}{2}g\mu_BB,
\end{equation}
where $g$ is the exciton $g$-factor and $\mu_B$ the Bohr magneton. We approximate the diamagnetic shift with a parabola with prefactor $\gamma$ \cite{VanHattem2013,Tholen2016}. The fine structure splitting of the studied QDs \cite{Huo2014} is negligibly small in this context. For the diamagnetic shift, a probe of the lateral area of the exciton, we find values in the range $\gamma=15-20\ \mu\text{eV}/\text{T}^2$, in good agreement with Ref.\ \onlinecite{Ulhaq2016}. The dependence of $\gamma$ on the VC-area ($A_{\text{VC}}$) is shown in Fig.\ \ref{fig:corr}(b). We find a correlation of $\rho=0.805$ between $A_{\text{VC}}$ and $\gamma$ which reveals a connection between the capture-zone area and the lateral size of a QD. An Al-droplet with a larger VC-area has a larger lateral extent leading to a QD with a weaker lateral confinement potential. This finding is also fully compatible with the capture-zone model. For the direct dependence between the emission energy, $E_{X^0}$, and the diamagnetic shift, $\gamma$, we find an approximately linear relation (Fig.\ \ref{fig:corr}(c)) associated with a correlation of $\rho=-0.922$. This connection between vertical and lateral confinement is consistent with a reported correlation between emission energy and $s$-to-$p$-shell--splitting \cite{Wang2009}.

A plot of the exciton $g$-factor versus $A_{\text{VC}}$ is shown in Fig.\ \ref{fig:corr}(d). The $g$-factor shows a slight dependence on $A_{\text{VC}}$ with $\rho=-0.562$. For these QDs, the electron $g$-factor is very small such that the exciton $g$-factor is determined largely by the hole $g$-factor \cite{Ulhaq2016,Beguin2018}. The hole states are predominantly heavy hole in character. However, even a small admixture of light-hole states reduces the $g$-factor from the heavy hole limit by a large amount \cite{Ares2013,Watzinger2016}. This light-hole admixture is size-dependent which can explain the dependence of the $g$-factor on $A_{\text{VC}}$. However, the $g$-factor is more weakly correlated with $A_{\text{VC}}$ than the emission energy. We speculate that the hole $g$-factor is sensitive to the shape and not just the volume of the QD leading to a weaker correlation with $A_{\text{VC}}$.

Shown in Fig.\ \ref{fig:corr}(e) is the splitting between neutral and positively-charged exciton ($E_{X^0}-E_{X^+}$) as a function of $A_{\text{VC}}$. Using single-particle wavefunctions, $E_{X^0}-E_{X^+}$ can be associated with the term $E_{eh}-E_{hh}$, where $E_{hh}$, $E_{eh}$ are the direct Coulomb integrals between two holes, and an electron and a hole, respectively \cite{Warburton1998,Cheng2003}. Both terms decrease with increasing lateral size of the QD and hence with increasing size of the VC-area, $A_{\text{VC}}$. Experimentally, we observe a monotonic relation between $A_{\text{VC}}$ and $E_{X^0}-E_{X^+}$ with a negative correlation ($\rho=-0.819$). For the direct relation between $E_{X^0}-E_{X^+}$ and the emission energy $E_{X^0}$, we find a linear dependence corresponding to a pronounced correlation of $\rho=0.942$ \cite{Dalgarno2008} (see Fig.\ \ref{fig:corr}(f)). This dependence also indicates a connection between the lateral and vertical confinements.

\textit{Conclusions} -- We show how the optical properties of QDs grown by GaAs-infilling of Al-droplet-etched nano-holes are connected to the capture-zone model, a concept from nucleation theory. The QD-positions and the optical QD-properties are obtained simultaneously by spatially resolved photoluminescence. The capture zone of QDs is determined by the Voronoi-diagram of the QD-positions. We find a strong negative correlation between the VC-area and the emission energy of QDs. This result can be explained with the capture-zone model applied to the growth-phase of the Al-droplets. A relation between VC-area and further optical properties (diamagnetic shift and neutral-to-charged exciton splitting) shows that not only the vertical but also the lateral QD-size is correlated with the area of the capture zone. We measure these correlations on samples with low QD-densities ($n_{\text{QD}}<1\ \mu\text{m}^{-2}$). Properties of a QD on a nm-scale, which determine its optical emission, are therefore connected to its surroundings on a $\mu$m-scale. This result might be transferable to other nanostructures when strong material diffusion is present during the growth. The correlations between different QD-parameters facilitate preselection of QDs for applications which place stringent requirements on the QD-properties. The correlation between emission energy and capture-zone area has a powerful implication: If all capture-zone areas are identical -- e.g.\ by forcing the nucleation of the Al-droplets on a lattice -- then a spectrally narrow wavelength distribution of the QD-ensemble can potentially be engineered. This idea has been successfully applied to stacked QD-layers and QDs in pyramidal nanostructures \cite{Teichert1996,Springholz1998,Pinczolits1999,Gogneau2004,Kiravittaya2006,Biasiol2011,Mohan2010}. For the system investigated here, Fig.\ \ref{fig:VC}(f) indicates that the QD-ensemble would narrow by a factor of two if all Voronoi-cell areas were in a range of $3-4\ \mu\text{m}^2$.

The authors thank Immo S\"ollner for fruitful discussions on the experiments and Christoph Kl\"offel for fruitful discussions on the QD hole states. MCL, LZ, JPJ, and RJW acknowledge financial support from SNF Project No.\ 200020\_156637 and from NCCR QSIT. LZ received funding from the European Union Horizon 2020 Research and Innovation programme under the Marie Sk\l{}odowska-Curie grant agreement No.\ 721394 (4PHOTON). AL, JR, and ADW gratefully acknowledge financial support from grants DFH/UFA CDFA05-06, DFG TRR160, and BMBF Q.Link.X 16KIS0867. AL acknowledges funding from DFG via project No. 383065199, LU2051/1-1. YH was supported by NSFC under grant No.\ 11774326, Shanghai Pujiang R\&D Program, and SHSF under grant No.\ 17ZR1443900. AR acknowledges support from the FWF P29603.

MCL, LZ, and JPJ carried out the experiments. MCL, LZ, AR, and RJW analysed the data. YH and AR performed the AFM-measurement. AR, YH, and OGS grew sample A; JR, ADW, and AL grew sample B. MCL initiated the project. MCL and RJW wrote the paper with input from all authors.

\bibliography{main_arxiv_v3.bbl}

\setcounter{equation}{0}
\renewcommand{\theequation}{S\arabic{equation}}
\renewcommand{\theHequation}{S.\theequation}

\setcounter{figure}{0}
\renewcommand{\thefigure}{S\arabic{figure}}
\renewcommand{\theHfigure}{S.\thefigure}

\onecolumngrid
\newpage

\section{Supplemental Material}
\subsection{Frequency Distribution of the Voronoi Cell Areas}
\label{sec:distr}
\begin{figure}[h]
\includegraphics[width=0.45\columnwidth]{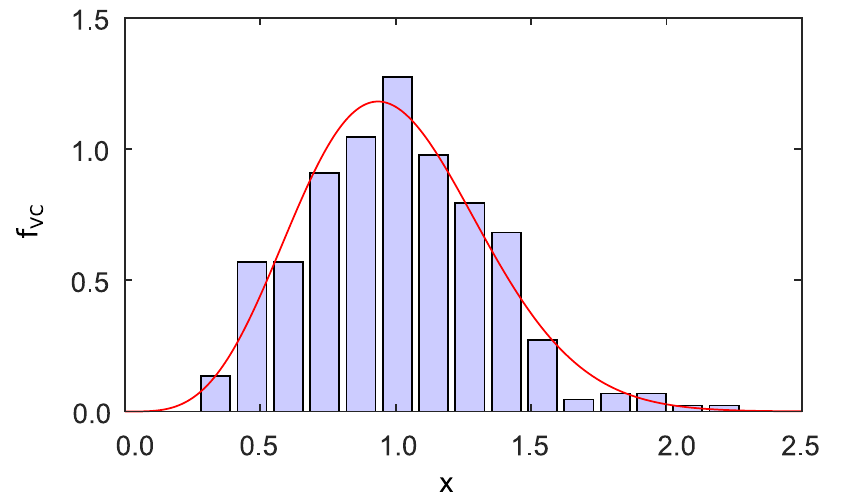}
\caption{\label{fig:VCdistribution}Frequency distribution, $f_{\text{VC}}$, of the normalized Voronoi-cell area, $A_{\text{VC}}/\langle A_{\text{VC}}\rangle\equiv x$. For better visibility, the normalized Voronoi-cell areas are divided in finite intervals. The red line is a fit of Eq.\ \ref{Eq:GWD} obtained by likelihood optimization.}
\end{figure}
The frequency distribution of the VC-areas gives further information about the Al-droplet formation. The probability density distribution, $f_{\text{VC}}$, of the normalized Voronoi cell areas, $A_{\text{VC}}/\langle A_{\text{VC}}\rangle\equiv x$, is often modeled by a generalized Wigner distribution \cite{Pimpinelli2007,Groce2012}:
\begin{equation}
f_{\text{VC}}(x\mid\eta)=a_{\eta}\cdot x^{\eta}\cdot\exp\left(-b_{\eta}x^2\right).\label{Eq:GWD}
\end{equation}
In our notation, $f(x\mid \eta)$ corresponds to a probability density distribution for $x$ under the condition $\eta$. The parameters $b_{\eta}$ and $a_{\eta}$ are defined by the constraint that $f_{\text{VC}}$ is a normalized probability density distribution with a mean of 1. Explicit expressions for $b_{\eta}$, $a_{\eta}$ are given in Refs.\ \onlinecite{Gonzalez2011,Einstein2014}. The parameter $\eta$ can be connected to the critical nucleus size, $i$, via $\eta\approx i+2\ $\cite{Groce2012,Einstein2014,Pimpinelli2014}.

We fit Eq.\ \ref{Eq:GWD} to the distribution of normalized VC-areas via likelihood optimization, without making approximations such as Poissonian or Gaussian error estimation \cite{Rossi2018}. Let $\{A_{\text{VC}}^{(i)}\}_{i\in\{1..N\}}$ be the set of the areas corresponding to the $N$ different Voronoi cells. We define $\{x^{(i)}\}_{i\in\{1..N\}}$ as the set of all normalized Voronoi cell areas, $x^{(i)}=A_{\text{VC}}^{(i)}/\langle A_{\text{VC}}\rangle$. Eq.\ \ref{Eq:GWD} assigns a probability density to each value $x^{(i)}$. We assume that every value $x^{(i)}$ corresponds to an independent random variable, $X^{(i)}$. Then, under the condition of a fixed value for $\eta$, the probability density of measuring a set of normalized Voronoi cells, $\{x^{(i)}\}_{i\in\{1..N\}}$, is given by:
\begin{equation}
\label{Eq:likelihood}
P\left(\{x^{(i)}\}_{i\in\{1..N\}}\mid\eta\right)=\prod_{i=1}^N f_{\text{VC}}(x^{(i)}\mid\eta).
\end{equation}
This likelihood distribution, $P\left(\{x^{(i)}\}_{i\in\{1..N\}}\mid\eta\right)$, is maximum for $\eta=\eta_{opt}=3.66$. The mean of the likelihood distribution is $\eta_m=3.68$. The difference between mean and maximum of the likelihood distribution is small because the distribution is close-to symmetric and only slightly biased. The found value of the parameter $\eta$ corresponds to a critical nucleus size of $i\approx2$.

The likelihood distribution, $P\left(\{x^{(i)}\}_{i\in\{1..N\}}\mid\eta\right)$, has a standard deviation of $\sigma_{\eta}=0.33$ when varying $\eta$ for a fixed measurement, $\{x^{(i)}\}_{i\in\{1..N\}}$. Bayes' theorem connects the probability density for measuring $\{x^{(i)}\}_{i\in\{1..N\}}$ under the condition of a fixed parameter $\eta$ to the probability density for $\eta$ under the condition of a measurement, $\{x^{(i)}\}_{i\in\{1..N\}}$:
\begin{equation}
\label{Eq:bayes}
P\left(\eta\mid\{x^{(i)}\}_{i\in\{1..N\}}\right)=\frac{P\left(\{x^{(i)}\}_{i\in\{1..N\}}\mid\eta\right)\cdot P\left(\eta\right)}{P\left(\{x^{(i)}\}_{i\in\{1..N\}}\right)}.
\end{equation}
The distribution $P\left(\{x^{(i)}\}_{i\in\{1..N\}}\right)$ does not depend on $\eta$. Furthermore, we assume a uniform prior distribution, $P\left(\eta\right)$. In this case, both probability density distributions $P\left(\eta\mid\{x^{(i)}\}_{i\in\{1..N\}}\right)$ and $P\left(\{x^{(i)}\}_{i\in\{1..N\}}\mid\eta\right)$ are equal up to a prefactor which does not depend on $\eta$. Therefore, the width of both distributions is identical. The error on the parameter $\eta$ is given by $\sigma_{\eta}=0.33$, the standard deviation calculated for the likelihood distribution, $P\left(\{x^{(i)}\}_{i\in\{1..N\}}\mid\eta\right)$.

Shown in Fig.\ \ref{fig:VCdistribution} is the distribution of the relative frequency, $f_{\text{VC}}$, of the normalized VC-area, $A_{\text{VC}}/\langle A_{\text{VC}}\rangle\equiv x$. In this figure, the normalized Voronoi cell areas, $\{x^{(i)}\}_{i\in\{1..N\}}$, are divided in finite intervals. The solid red line is the fit ($\eta=\eta_{opt}=3.66$) of Eq.\ \ref{Eq:GWD} to the data, $\{x^{(i)}\}_{i\in\{1..N\}}$.

\subsection{Correction of the Non-Linearity of the Piezo-Scanners}
\begin{figure}[h]
\includegraphics[width=1\columnwidth]{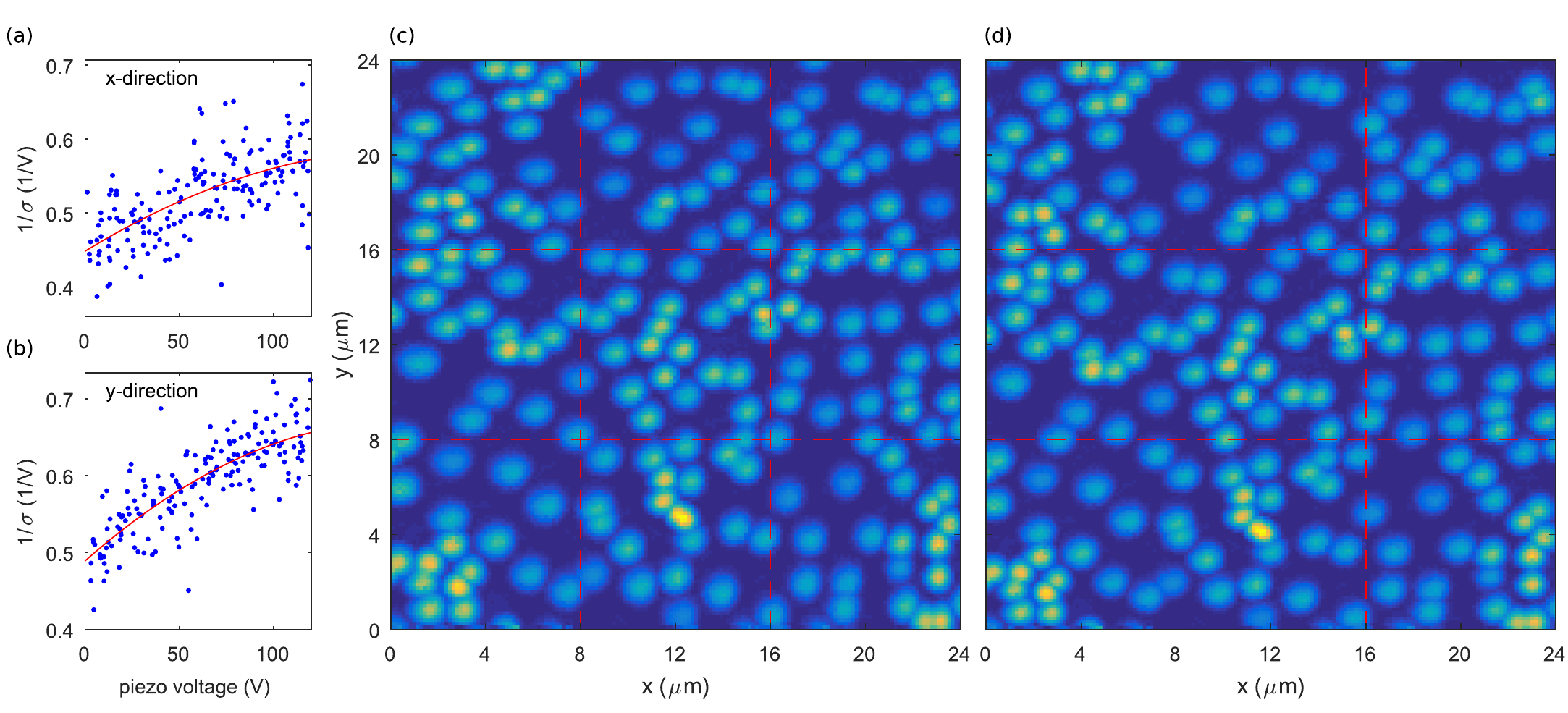}
\caption{\label{fig:distCorr}Correction of the distortion due to the non-linearity of the piezo-scanners. The QD-position is set by the voltages $V_{x/y}$ applied to the piezo-scanners. The distortion correction is carried out independently for the (a) x- and (b) y-directions (x-direction in the following explanation). Initially, the spot size $\sigma_{x}$ is measured for every QD in units of the voltage applied to the piezo-scanner. It is obtained by fitting a 2D Gaussian to the PL-intensity which is spectrally filtered to select individual QDs. For a larger derivative $\frac{\text{d}x}{\text{d}V_x}$, the spot size appears smaller than for a smaller value of $\frac{\text{d}x}{\text{d}V_x}$. The spot size $\sigma_x$ is therefore inversely proportional to the derivative $\frac{\text{d}x}{\text{d}V_x}$. The red curves in (a), (b) are fits to a phenomenological parabolic dependence between applied piezo-voltage $V_x$ and $1/\sigma_x$: $\frac{\text{d}x}{\text{d}V_x}=\tilde{c}/\sigma_x=\tilde{c}(a_0+a_1V_x+a_2V_x^2)$. We use the fit results to map $V_x$ to the position $x(V_x)=\tilde{c}\int_0^{V_x}(a_0+a_1V+a_2V^2)\ \text{dV}$. The prefactor $\tilde{c}$ is obtained by the constraint that the highest voltage corresponds to the full scan range. (c) PL-intensity as a function of the xy-position. The PL is integrated over the full QD-ensemble. The PL-map is shown without the distortion correction assuming that position and applied piezo-voltage are related linearly. Due to the described non-linearity of the piezo-scanners, the QD-spots at low voltages appear slightly larger in comparison to the QD-spots at high voltages -- the PL-map is distorted. (d) The same PL-map with a distortion correction using the above method. The QD-spot sizes are homogeneous indicating a successful correction of the distortion.} 
\end{figure}

The determination of Voronoi-cell sizes is based on measuring the PL as a function of the position. Such a PL-map is carried out by scanning the sample position with piezoelectric xy-scanners. The scanners have a hysteresis which, however, does not affect our measurements as we perform all measurements while scanning in the forward direction. Besides, the piezo-scanner position depends non-linearly on the applied voltage. This non-linearity could potentially be corrected by calibrating the piezo-scanners with a well-defined reference-structure. Here, we use a different approach: The emission spot of a QD appears differently in size depending on the absolute position of the scanner. The QD spot size is a direct measure of the non-linear dependence between the applied piezo-voltage and the position. The lateral size of a QD itself ($<40\ \text{nm}$) is negligible in comparison to the spot size. We can, therefore, use the widths of the QD spots as a reference to compensate for the distortion of the PL-map. The corresponding procedure is explained in the caption of Fig.\ \ref{fig:distCorr}. The distortion correction works well due to the large number of QDs on each PL-map. A comparison of a PL-map with and without distortion correction is shown in Fig.\ \ref{fig:distCorr}(c, d).
\label{sec:piezo}

\newpage
\subsection{AFM-Measurement}
\label{sec:AFM}
\begin{figure}[h]
\includegraphics[width=0.9\columnwidth]{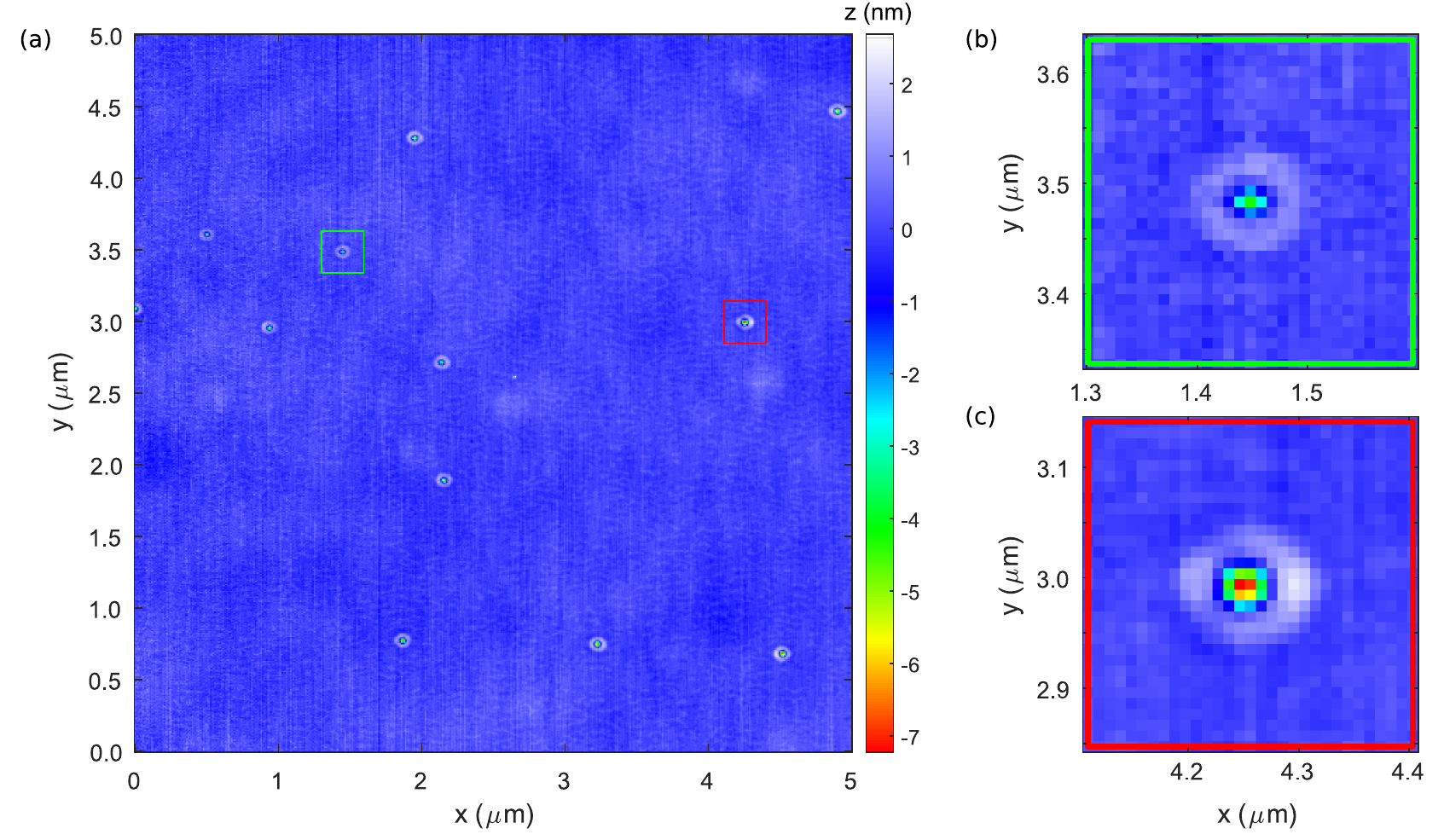}
\caption{\label{fig:AFMcorr}(a) An atomic force microscopy (AFM) image on a sample similar to sample A. The growth is stopped after etching the nano-holes. The size of the AFM-scan is $5\times5\ \mu m^2$ with $512\times512$ pixels. The AFM-image indicates that the more separated nano-holes (with a larger VC-area) are deeper than those with a close-by neighbor. This finding is in good agreement with the results shown in the main text. A zoom-in of two nano-holes is shown in (b, c) to illustrate this observation. The first nano-hole has several close-by neighbors and is shallow. In contrast, the second nano-hole is more isolated and is particularly deep. The image size and resolution are too low to allow for a quantitative statistical analysis comparable to the main text. Note that an AFM-image comparable in size to the presented PL-measurements, simultaneously imaging individual nano-holes with $\sim2\ \text{nm}$ resolution, would have to be $~\sim10^4\times10^4$ pixels large -- a very time-consuming measurement.
}
\end{figure}

We perform an AFM-measurement on a reference sample for which the growth is stopped after etching of the nano-holes. An AFM-image of a $5\times5\ \mu m^2$  large region is shown in Fig.\ \ref{fig:AFMcorr}. The measurement suggests that nano-holes with other nano-holes in their proximity are shallower in comparison to more separated ones. This finding is consistent with the capture zone model and the results obtained by spatially resolved photoluminescence. The size of the AFM-image does not allow for a quantitative analysis comparable to which is done based on the spatially resolved PL-measurements.

\newpage
\subsection{Emission Energy vs. Voronoi Cell Area}
\label{sec:EVCcorr}
\begin{figure}[h]
\includegraphics[width=0.5756\columnwidth]{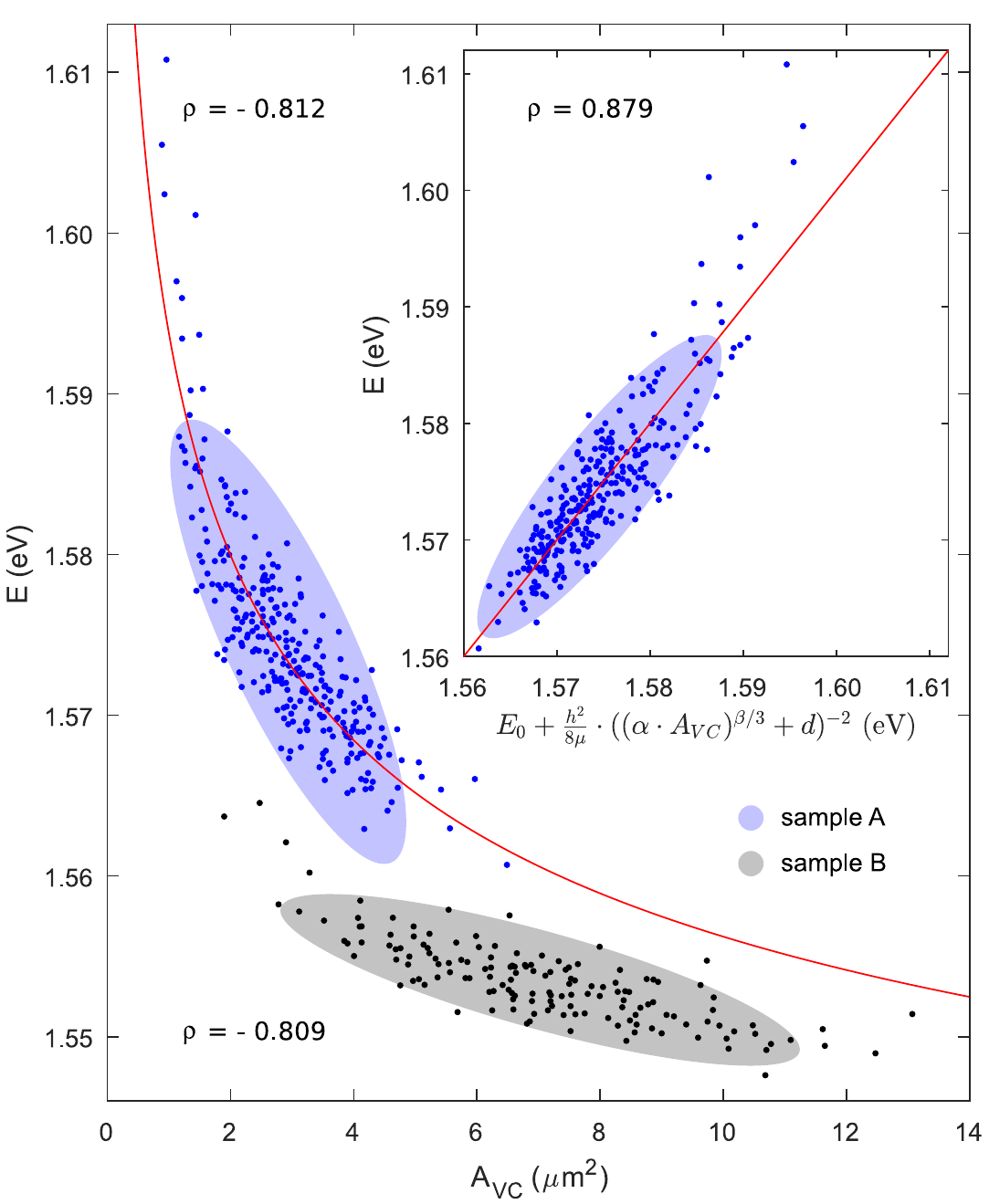}
\caption{\label{fig:Eq1}Emission energy ($X^0$) as a function of Voronoi-cell area ($A_{\text{VC}}$). The red line is a fit to Eq.\ 1 of the main text for QDs on sample A (blue points). Inset: emission energy of QDs on sample A as a function of the expression on the right-hand side of Eq.\ 1 of the main text. The relation between both quantities is more linear (correlation coefficient $\abs{\rho}=0.879$) than the direct relation of emission energy and $A_{\text{VC}}$ ($\abs{\rho}=0.812$) which supports Eq.\ 1 of the main text. QDs on sample B (black data points) are red-shifted compared to QDs on sample A.}
\end{figure}
Shown in Fig.\ \ref{fig:Eq1} is the dependence between the emission energy and the Voronoi cell (VC) area, $A_{\text{VC}}$, for the two different samples (A, B). For both samples, we observe that the emission energy decreases with increasing Voronoi-cell area, $A_{\text{VC}}$.

Eq.\ 1 of the main text suggests that the relation between $A_{\text{VC}}$ and emission energy is non-linear. In Fig.\ \ref{fig:Eq1} we plot the emission energy as a function of the expression on the right-hand side of Eq.\ 1. The correlation of this dependence is higher ($\abs{\rho}=0.879$) than the correlation between emission energy and $A_{\text{VC}}$ ($\abs{\rho}=0.812$). This result suggests that Eq.\ 1 of the main text is a better description of the data than a linear approximation.

\end{document}